\documentstyle[psfig,12pt]{article}

\newcommand{\CE}{{\cal E}}

\newcommand{\lesssim}{ \mathop{}_{\textstyle \sim}^{\textstyle <} }
\newcommand{\bea}{\begin{eqnarray}}  \newcommand{\eea}{\end{eqnarray}}
\newcommand{\beq}{\begin{equation}}  \newcommand{\eeq}{\end{equation}}
\newcommand{\non}{\nonumber}  
  
\newcommand{\lmk}{\left(}  \newcommand{\rmk}{\right)}

\newcommand{\del}{\partial}  

\newcommand{\bib}{\bibitem} 
\newcommand{\la}{\left\langle} \newcommand{\ra}{\right\rangle}
 
\newcommand{\gtilde} {~ \raisebox{-1ex}{$\stackrel{\textstyle >}{\sim}$} ~} 
\newcommand{\ltilde} {~ \raisebox{-1ex}{$\stackrel{\textstyle <}{\sim}$} ~}

\newcommand{\VS}{\la S \ra}
\newcommand{\VFS}{\la F_S \ra}
\newcommand{\gm}{m_{3/2}}
\newcommand{\mm}{m_{\phi}}
\newcommand{\GM}{\lmk \frac{m_{3/2}}{100 {\rm keV}} \rmk}

\newcommand{\SS}{\lmk \frac{\VS}{10^{5} {\rm GeV}} \rmk}
\newcommand{\LL}{\lmk \frac{\Lambda}{10^{4} {\rm GeV}} \rmk}
\newcommand{\g}{\lmk \frac{g_{\ast}}{200} \rmk}
\newcommand{\G}{\,{\rm GeV}}

\begin{document}


\begin{center}
{\Large \bf Cosmological Moduli Problem and Oscillating Inflation
  in Gauge-Mediated Supersymmetry Breaking \\}
\vskip 1cm
{\large T. Asaka} \\
\vskip 0.2cm
{\large \em Institute for Cosmic Ray Research,
  University of Tokyo, \\
  Tanashi, 188-8502, Japan}
\vskip 0.5cm {\large M. Kawasaki} \\ 
\vskip 0.2cm {\large \em Institute for Cosmic Ray Research,
  University of Tokyo, \\
  Tanashi, 188-8502, Japan}
\vskip 0.5cm {\large Masahide Yamaguchi} \\ 
\vskip 0.2cm {\large \em Department of Physics,
  University of Tokyo, \\
  Tokyo, 113-0033, Japan}
\vskip 0.5cm
{\large October, 13, 1998}
\end{center}


\begin{abstract}
    We investigate cosmological moduli problem in the gauge-mediated
    supersymmetry breaking (GMSB). A mini-inflation (oscillating
    inflation) takes place when a scalar field corresponding to the
    flat direction in GMSB oscillates along the logarithmic potential
    induced by the gauge-mediation mechanism. It is shown that this
    oscillating inflation can sufficiently dilute the relic abundance
    of the string moduli for some ranges of parameters in the GMSB
    models.
\end{abstract}

\thispagestyle{empty} \setcounter{page}{0} \newpage
\setcounter{page}{1}

\section{Introduction}

\label{sec:introduction}

\indent

Supersymmetry (SUSY) is an attractive candidate beyond the standard
model because it stabilizes the electroweak scale against the
radiative corrections. However, SUSY is not exact symmetry and should
be broken since we have not detected degenerate superpartners of
ordinary particles. Dynamics of the SUSY breaking has not been clear
and been one of the most important open questions left to us.  So far
many ideas have been proposed as the SUSY breaking mechanism.  Among
them the gauge-mediated SUSY breaking (GMSB) mechanism\cite{GMSB} is
very attractive since it gives a solution of the SUSY flavor problem.

However, the models of GMSB are faced with a serious cosmological 
problem.  One of the consequences of GMSB is the existence of a light 
stable gravitino.  Such a light gravitino is produced after the 
primordial inflation and if its mass is $m_{3/2} \gtilde 1$ keV, the 
reheating temperature of the inflation should be low enough \cite{MMY}.

Furthermore, when one considers GMSB in the framework of the
superstring theories, more severe cosmological difficulty,
``cosmological moduli problem'', arises.  As a general consequence of
the superstring theories, light moduli particles $\phi$ appear and
their masses $m_\phi$ are expected of the order of the gravitino
mass~\cite{mp}.  With such small masses $m_\phi \simeq m_{3/2}$, since
they have only the gravitationally suppressed interaction, they are
stable within the age of our universe\,($\sim 10^{17}$~sec) and easily
overclose the present critical density of the universe.

The thermal inflation model proposed by Lyth and Stewart~\cite{LS}
gives a solution of this problem.  In fact it was shown in
Ref.~\cite{GMM} that the above overclosure problem could be solved if
one assumed the thermal inflation.  However, it was pointed out
\cite{KY} that if $m_\phi \gtilde 100$ keV, a more stringent
constraint comes from the observation of the present cosmic
x($\gamma$)-ray backgrounds, which excludes the moduli with mass 100
keV $\ltilde m_\phi \ltilde $ 1 GeV even if the universe
experienced the thermal inflation \cite{HKY,AHKY}.

Recently, Moroi\,\cite{Moroi} reported that the SUSY breaking field,
which has a very flat potential, gives another dilution mechanism of
the relic abundance of the moduli.  This is because an accelerating
universe (say an oscillating inflation\,\cite{OI}) takes place
when the scalar field corresponding to the flat direction oscillates
along the logarithmic potential induced by the GMSB
mechanism. In Ref.\cite{Moroi} the specific model of GMSB was
analyzed and was shown that the moduli is significantly diluted.
However, during the oscillating inflation, the minimum of the
moduli potential deviates from the true minimum through the additional
SUSY breaking effect due to the large vacuum energy of the inflaton
field.  After the inflation ends, the moduli field begins to move
toward the true minimum and hence the (secondary) oscillation starts
even if the amplitude of the original coherent oscillation of the
moduli sufficiently decreases during the inflation.  However, this
deviation of the minimum was not properly taken into account and the
moduli density coming from the secondary oscillation was totally
neglected in Ref.~\cite{Moroi}.

In this letter we consider the cosmological moduli problem assuming
that the oscillating inflation originates in the minimal SUSY standard
model (MSSM) flat direction\footnote{Our discussion applies to a
general flat direction of the GMSB models.} and taking account of the
secondary oscillation. In the following, we discuss the oscillating
inflation which is a general consequence of the flat directions in the
GMSB models and examine how dilution mechanism works.  Then we study
whether this oscillating inflation could solve the cosmological
difficulties of the moduli or not.  Finally, we summarize our
conclusions.

\section{Oscillating Inflation and Entropy Production}

\label{sec:oscillating}

\indent

In the generic SUSY models, there appears the flat direction, say $X$,
in the scalar potential.  However, the effects of SUSY breaking lift
the flat direction.  In most models of GMSB, SUSY is broken in the
dynamical SUSY-breaking sector and its effects are transferred into
the messenger sector.  The messenger sector contains a gauge singlet
multiplet $S$ which is supposed to have a $A$-component vacuum
expectation value (vev) $\VS$ and also have a $F$-component vev
$\VFS$. Then SUSY is broken in the messenger sector and its effects
are mediated to the ordinary sector through the standard model (SM)
gauge interaction by integrating out the heavy messenger fields (for a
review see Ref.~\cite{GM}).  For example, the sfermions
$\widetilde{f}$ acquire masses of the order of
\begin{eqnarray}
    m^2_{\widetilde{f}} \simeq 
    \left( \frac{ \alpha_{SM} }{ 4 \pi } \right)^{2} \Lambda^{2},
\end{eqnarray}
where $\alpha_{SM}$ denotes the appropriate coupling of the SM gauge 
interaction, and $\Lambda$ is a ratio between $\VFS$ and $\VS$ and 
should be $\Lambda = \VFS / \VS \gtilde$ 10 TeV from the present 
experimental limits on the masses of the superparticles.

The flat direction $X$ obtains a potential through the gauge
meditation mechanism. For a large $X$ ($|X| \gg \VS$), the potential
is expressed as (e.g. see Ref.\cite{GMM})
\begin{eqnarray}
    \label{V-Gauge}
    V(X) \simeq 
    \left( \frac{ \alpha_{SM} }{ 4 \pi } \right)^{2} 
    \VFS^2 
    \left( \ln \frac{ |X|^2 }{ \VS^2 } \right)^2
    + \cdots.
\end{eqnarray}
Here we have assumed the positive sign of overall factor.  However, 
for an extremely large $X$, we cannot neglect the contribution from 
the supergravity.  The SUSY breaking effects communicated by the 
gravity are expected to give the soft mass to the flat direction $X$ 
which is of the order of the gravitino mass $m_{3/2}$.  Then the 
potential of $X$ due to the gravity is expressed as
\begin{eqnarray}
    \label{V-Gravity}
    V(X) \simeq m_{3/2}^2 |X|^2 + \cdots.
\end{eqnarray}
Comparing Eq.(\ref{V-Gauge}) with Eq.(\ref{V-Gravity}), the 
contribution from the gravity-mediation effect becomes more important 
for the region $|X| \gtilde X_{eq}$, where $X_{eq}$ is estimated as
\begin{eqnarray}
    \label{X_eq}
    X_{eq} &\simeq& \frac{ \VFS }{ m_{3/2} } \nonumber \\
    &\simeq& 10^{13}\,\G \SS \LL \GM^{-1} \:.
\end{eqnarray}
Here note that in Eq.(\ref{V-Gauge}) the logarithmic factor
and the gauge coupling constants are almost cancelled.%
\footnote{
  In the following analysis, we will drop off such an order one
  constant to make a more conservative analysis.}
In order that $X_{eq}$ be smaller than about the Planck scale $M_G =
2.4 \times 10^{18}$ GeV, $\VS$ should be constrained as%
\footnote{
  This upper bound on $\VS$ ensures that the $F$-vev in the dynamical
  SUSY breaking sector $\la F_{DSB} \ra$ is equal to or larger than
  $\VFS$, i.e., $\la F_{DSB} \ra \sim m_{3/2} M_G \gtilde \VFS
  \sim \VS \Lambda$. }
\begin{eqnarray}
    \label{C-VS}
    \VS &\ltilde& \frac{ M_G m_{3/2} }{ \Lambda } \nonumber \\
        &\simeq& 2.4 \times 10^{10} ~\mbox{GeV} ~
        \left( \frac{ m_{3/2} }{ 100~\mbox{keV} } \right)
        \left( \frac{ \Lambda }{ 10^{4}~\mbox{GeV} } \right)^{-1}.
\end{eqnarray}

Then we turn to discuss the cosmological evolution of the flat
direction $X$ in the inflationary universe.  The initial value of $X$
when the primordial inflation ends ($|X| = X_0$) is crucial in this
discussion.  Our claim here is that the $X_0$ is much larger than
$\VS$ because of the chaotic condition of the early universe.%
\footnote{
The additional SUSY breaking effects during the primordial inflation
may naturally explain the initial condition $X_0$. }

First, we consider the case that $X_0 \gtilde X_{eq} (\gg \VS)$.
Since the effective mass of $X$ after the primordial inflation is of
the order of the gravitino mass, $X$ starts to roll down along the
potential ($\ref{V-Gravity}$) when the Hubble parameter, $H$, of the
universe becomes nearly equal to $m_{3/2}$.  At this time, the cosmic
temperature of the universe $T_{3/2}$ is estimated as
\begin{eqnarray}
    \label{T32}
    T_{3/2} &\simeq& 1.7 g_{\ast}^{-1/4} \sqrt{M_{G}\gm} 
    \nonumber \\
           &\simeq& 7.2 \times 10^6 \G \g^{-\frac14} \GM^{\frac12} \:,
\end{eqnarray}
where $g_\ast$ is the effective degree of the relativistic freedom.
Here we have assumed that the reheating process of the primordial
inflation had been completed at $T > T_{3/2}$. For $T \ltilde
T_{3/2}$, the flat direction $X$ causes a coherent oscillation with
the initial amplitude $X_0$.  As the universe expands, the amplitude
decreases as $|X| \propto R^{-3/2}$ ($R$: the scale factor) due to the
parabolic form of the potential (\ref{V-Gravity}).

When the amplitude of the oscillation becomes smaller than $X_{eq}$ 
[Eq.(\ref{X_eq})], the cosmological evolution of $X$ is drastically changed 
since the logarithmic potential (\ref{V-Gauge}) governs the dynamics 
in this region.  The temperature at $|X| \simeq X_{eq}$ is given by
\begin{eqnarray}
    \label{T_eq}
    T_{eq} \simeq T_{3/2} 
    \left( \frac{X_{eq} }{ X_0 } \right)^{\frac23}.
\end{eqnarray}
The evolution of $X$ is understood as follows; from the virial
theorem, we have a relation,
\beq
  2\la K \ra = \la \frac{\del V}{\del X}X 
        + \frac{\del V}{\del X^{\ast}} X^{\ast} \ra \:,
\eeq  
where $K$ is the kinetic energy of $X$ and the bracket represents time
average over a cycle.  This leads to $\la K \ra = \la 2V /
\ln(|X|^2/\VS^2) \ra$ for $|X| \ltilde X_{eq}$ and $\la K \ra = \la V
\ra$ for $|X| \gtilde X_{eq}$. With the help of the equation of motion
of $X$, the total energy density, $\CE \equiv K + V$, behaves as
$d\CE/dt = -6H K$ and the amplitude $X$ decreases as $|X|
\propto R^{-3}$ for $|X| \ltilde X_{eq}$ and $|X|^2 \propto R^{-3}$
for $|X| \gtilde X_{eq}$. Thus, the energy density of a flat direction
rapidly dominates the universe under the logarithmic potential.%
\footnote{This situation does not change even if there exists a
coherent oscillation of the moduli field.} It should be noted that an
accelerating universe does occur in spite of the oscillation of the
field \cite{OI}. To confirm this, we define an averaged adiabatic
index, $\gamma$, as
\bea
  \gamma &\equiv& \la \frac{2K}{K + V} \ra \non \\
         &=& \frac{4}{\ln \frac{|X|^2}{\VS^2} +2} \:,
\eea
For a successful inflation we need $\gamma < 2/3$, that is, $\ln
\frac{|X|^2}{\VS^2} > 4$.  Thus, till the amplitude of $X$ becomes of
the order of $\VS$, the inflation due to the logarithmic potential
(\ref{V-Gauge}) takes place.  We call this inflation ``oscillating
inflation''.

When the oscillating inflation ends at $|X| \sim \VS$,
the temperature of the universe is estimated as
\bea
\label{TS}
  T_{\rm S} &\simeq& T_{eq} \lmk \frac{\VS}{X_{eq}} \rmk^{\frac13} 
             \non \\
        &\simeq& T_{3/2}
                     \lmk \frac{X_{eq}}{X_{0}} \rmk^{\frac23}
                     \lmk \frac{\VS}{X_{eq}} \rmk^{\frac13} 
             \non \\
        &\simeq& 3.9 ~\mbox{GeV} 
        \left( \frac{ g_\ast }{ 200 } \right)^{-\frac14} 
        \left( \frac{ m_{3/2} }{ 100 ~\mbox{keV} } \right)^{\frac16} 
        \left( \frac{ \VS }{ 10^5~\mbox{GeV} } \right)^{\frac23}
        \left( \frac{ \Lambda }{ 10^4 ~\mbox{GeV} } \right)^{\frac13}
        \left( \frac{ X_{0} }{ M_{G} } \right)^{-\frac23} \:.
\eea
After the oscillating inflation it is expected that the flat direction
$X$ causes an oscillation around its true minimum $\la X \ra
\ltilde \VS$ and the $X$ decay occurs when the Hubble parameter
becomes of the order of its decay width. At this epoch, the vacuum
energy $V_0$ of the oscillating inflation is transferred into the
thermal bath of the universe and reheats the universe to $T = T_{RX}$.
Here we do not specify the explicit dynamics during this reheating
epoch, but simply estimate the entropy production rate considering the
reheating temperature, $T_{RX}$, as a free parameter.%
\footnote{
  In fact, one example of such a low reheating process after the
  oscillating inflation is explained in Ref.\cite{Moroi}.}

If the vacuum energy $V_0$ is completely transferred into the thermal
bath, the oscillating inflation can increase the entropy by a factor
$\Delta$:
\bea
\label{Delta}
    \Delta  &\simeq& \frac{ 4 }{ 3 }
    \frac{ V_0 }{ \frac{ 2 \pi^2 }{ 45 } g_\ast T_S^3 T_{RX} } \non \\
    &\simeq&
    0.15 \frac{ \Lambda M_G^{\frac12} }{ m_{3/2}^{\frac12} T_{RX} }
    \g^{-\frac14}
    \left( \frac{ X_0 }{ M_G } \right)^{2} \non \\
    &\simeq&
    2.4 \times 10^{16} \g^{-\frac14} 
    \left( \frac{ \Lambda }{ 10^4~\mbox{GeV} } \right)
    \left( \frac{ m_{3/2} }{ 100 ~\mbox{keV} } \right)^{-\frac12} 
    \left( \frac{ T_{RX} }{ 10 ~\mbox{MeV} } \right)^{-1}
    \left( \frac{ X_0 }{ M_G } \right)^{2}.
\eea
Here we have used the fact that $V_0 \sim \VFS^2$.
\footnote{
Here we have dropped off the gauge coupling in 
Eq.(\ref{V-Gauge}) to make a conservative analysis.}
From Eq.(\ref{Delta}) we find that the maximum entropy production rate
is achieved when we take the lowest value of the reheating temperature
and the largest value of the initial displacement of $X$\,($X_0 \simeq
M_G$). To keep the success of the big-bang nucleosynthesis, we require
that the reheating temperature should be larger than $10$MeV.
Therefore, the oscillating inflation can dilute unwanted particles
which survive until late times of the thermal history of the universe.

To end this section, we briefly mention about the initial amplitude
$X_0$. We have assumed that $X_0$ might take an arbitrary value ($X_0
\gg \VS$) and considered the case $X_0 \gtilde X_{eq}$.  As shown in
Eq.(\ref{Delta}) we find that the maximum entropy production is
achieved when $X_0$ takes its maximum value of the order of the Planck
scale.  Thus we obtain less dilution factor for the case $X_0 \ltilde
X_{eq}$. In the following we assume $X_0 \simeq M_G$.

\section{Cosmological Moduli Problem with Oscillating Inflation}

\label{sec:abundance}

\indent

In this section, we assume the dilution mechanism of the 
oscillating inflation and examine whether it could solve the 
cosmological moduli problem or not.

The moduli $\phi$ starts to oscillate with the initial amplitude
$\phi_0 \sim M_G$ when its mass $m_\phi$ becomes comparable to the
Hubble parameter at $T = T_{\phi}$. Since the moduli mass is $m_\phi
\simeq m_{3/2}$, $T_\phi$ is almost same as $T_{3/2}$ [Eq.(\ref{T32})].  We
call this moduli ``big-bang moduli''.  At $T = T_\phi$, the ratio
of the energy density of this oscillating moduli to the entropy
density is given by~\cite{LS}
\bea
  \left( \frac{\rho_{\phi}}{s} \right)_{BB} 
      &\simeq&
           \frac{\frac12 m_{\phi}^2 \phi_{0}^2}
            {\frac{2\pi^2}{45} g_{\ast} T_{\phi}^3 }  \non \\
      &\simeq&
           9.0 \times 10^5 ~\mbox{GeV}~
           \left( \frac{ g_\ast }{ 200 } \right)^{-\frac14} 
           \left( \frac{ m_\phi }{ 100~\mbox{keV} } \right)^{\frac12}
           \left( \frac{\phi_{0}}{M_{G}} \right)^2 \:.
\eea
Since $\rho_{\phi} \propto R^{-3}$, this ratio takes a constant value
if no entropy is produced. Here it should be noted that the flat
direction $X$ and the moduli $\phi$ starts to oscillate at almost the
same time. Therefore the oscillating inflation always occurs after the
big-bang moduli oscillation begins and can dilute its abundance.
Using Eq.(\ref{Delta}), the relic abundance
of the big-bang moduli after the oscillating inflation is given by
\bea
    \label{RS-BB}
    \left( \frac{ \rho_\phi }{ s } \right)_{BB,0}
    &=&
    \left( \frac{ \rho_\phi }{ s } \right)_{BB} \times
    \frac{1}{\Delta}
    \nonumber \\
    &\simeq& 0.38 ~ \frac{ m_{3/2} T_{RX} }{ \Lambda }
    \left( \frac{ m_\phi }{ m_{3/2} } \right)^{\frac12}
    \left( \frac{ X_0 }{ M_G } \right)^{-2}
    \left( \frac{ \phi_0 }{ M_G } \right)^{2} \non \\
    &\simeq&
    3.4 \times 10^{-11} ~\mbox{GeV}~
    \left( \frac{ m_{3/2} }{ 100 ~\mbox{keV} } \right)
    \left( \frac{ \Lambda }{ 10^4 ~\mbox{GeV} } \right)^{-1}
    \left( \frac{ T_{RX} }{ 10 ~\mbox{MeV} } \right) \non \\
    &&~~~~~~~~~\times
    \left( \frac{ m_\phi }{ m_{3/2} } \right)^{\frac12}
    \left( \frac{ X_0 }{ M_G } \right)^{-2}
    \left( \frac{ \phi_0 }{ M_G } \right)^{2}.
\eea
You should notice that this abundance is independent of $\VS$.

There is another contribution to the moduli density other than the 
big-bang moduli.  During the oscillating inflation the moduli is 
displaced from its true minimum due to the additional SUSY breaking 
effects.  This displacement $\delta \phi$ is estimated as \cite{LS}
\begin{eqnarray}
    \delta \phi \sim \frac{ 3 H_{OI}^2 }{ m_\phi^2 + 3 H_{OI}^2 } \phi_0
    \sim \frac{ 3 H_{OI}^2 }{ m_\phi^2 } \phi_0,
\end{eqnarray}
where $H_{OI}$ is the Hubble parameter during the oscillating
inflation and $H_{OI} \ll m_\phi$.  When the oscillating inflation
ends at $|X| \sim \VS$, the moduli displacement from the true minimum
is $\delta \phi_0 \sim (V_0\phi_{0})/(m_\phi^2 M_G^2)$.  After this
inflationary epoch ($|X| \ltilde \VS$), since the logarithmic
potential is not effective, the moduli displacement goes to zero at
the rate $\delta \phi \propto R^{-3}$ as the universe expands.  On the
other hand, the amplitude of the big-bang moduli decreases at the rate
$\phi \propto R^{-3/2}$ and can not catch up with the motion of the
$\delta \phi$.  This causes another coherent oscillation of the moduli
\cite{LS},\footnote{ The occurrence of the secondary oscillation of the
  moduli can be easily understood from the point of view of the energy
  conservation.} which was totally neglected in Ref.~\cite{Moroi}.
                                                           
We call this moduli ``oscillating inflation moduli''.  The ratio of
the energy density of this oscillation to the entropy density is
estimated as
\begin{eqnarray}
    \left( \frac{ \rho_\phi }{ s } \right)_{OI}
    \simeq 
    \frac{ \frac{1}{2} m_\phi^2 \delta \phi_0^2 }
    { \frac{ 2 \pi^2 }{ 45 } g_\ast  T_S^3 }.
\end{eqnarray}
Then the present abundance of the oscillating inflation moduli 
is given by
\begin{eqnarray}
    \left( \frac{ \rho_\phi }{ s } \right)_{OI,0}
    &\simeq&
    0.38 ~ \frac{ \VS^2 \Lambda^2  T_{RX} }{m_{3/2}^2 M_G^2 }
    \left( \frac{ m_\phi }{ m_{3/2} } \right)^{-2}
    \left( \frac{ \phi_0 }{ M_G } \right)^{2} \non \\
    &\simeq&
    6.5 \times 10^{-14} ~\mbox{GeV}~
    \left( \frac{ m_{3/2} }{ 100 ~\mbox{keV} } \right)^{-2}
    \left( \frac{ \Lambda }{ 10^4 ~\mbox{GeV} } \right)^{2}
    \left( \frac{ \VS }{ 10^5 ~\mbox{GeV} } \right)^{2}
    \left( \frac{ T_{RX} }{ 10 ~\mbox{MeV} } \right)
    \nonumber \\
    &&~~~~~~~~~~~\times
    \left( \frac{ m_\phi }{ m_{3/2} } \right)^{-2}
    \left( \frac{ \phi_0 }{ M_G } \right)^{2}. 
\end{eqnarray}
Comparing with the abundance of the big-bang moduli~[Eq.(\ref{RS-BB})],
this abundance does depend on the value of $\VS$ and becomes important
for the lighter gravitino. Furthermore, both abundances take
minimum values for the lowest reheating temperature.  We show in Fig.1
and Fig.2 the present abundances of these two moduli with $T_{RX} =
10$~MeV in term of the density parameter defined by
\begin{eqnarray}
    \Omega_\phi h^2 \equiv \frac{ ( \rho_\phi /s )_0 }{ (\rho_c/s_0) },
\end{eqnarray}
where $( \rho_c / s_0 )$ is the ratio of the critical density to the
present entropy density:
\begin{eqnarray}
    \frac{\rho_{c}}{s_0} \simeq 3.6 \times 10^{-9} h^2 ~\mbox{GeV},
\end{eqnarray}
with $h$ the present Hubble constant in units of 100\,km/sec/Mpc.

Next we see the cosmological constraints on the moduli abundance.  
Since in the GMSB models the mass of the gravitino is predicted as 
$m_{3/2} (\simeq m_\phi) \ltilde 1$ GeV, the moduli decays most 
likely into two photons through the nonrenormalizable interaction 
suppressed by the gravitational scale.  Thus, the typical lifetime is 
estimated as \cite{KY,HKY}
\bea
  \tau_{\phi} &\simeq& \frac{64\pi}{b^2} \frac{M_{G}^2}{\mm^3} \non \\
              &\simeq& 7.6 \times 10^{23} \frac{1}{b^2} 
                        \lmk \frac{1 {\rm MeV}}{\mm} \rmk^3 {\rm sec}
                        \:.
\eea
where $b$ is a constant of order unity depending on the models of the
superstring theory.  Thus, the moduli becomes stable within the age of
the universe if $\mm \ltilde 100$\,MeV and it should not overclose
the universe\,($\Omega_\phi \ltilde 1$). We show this limit in Figs.1
and 2.

Moreover, as pointed out in Ref.\cite{KY} the moduli with mass $m_\phi
\gtilde 100$ keV is more stringently constrained from the observation
of the present x($\gamma$)-ray background spectrum since photons
produced by the moduli decay directly contribute to the spectrum.
This leads to the upper bound on $\Omega_\phi$ as shown in Figs.1 and
2. (Details are found in \cite{KY,HKY,AHKY}.)

From the figures it is seen that the abundance of the big-bang moduli
[Eq.(\ref{RS-BB})] puts the upper bound on $m_\phi \simeq m_{3/2}$.
This abundance is inversely proportional to $\Lambda$ and the upper
bound on $\Lambda \ltilde $ 100 TeV from the naturalness leads to
$m_\phi \ltilde 1$ MeV. On the other hand, the abundance of the
oscillating inflation moduli puts the lower bound on $m_\phi$.  Since
this abundance is proportional to $\VS$ and $\Lambda$, the minimum
values of them~($\VS > \Lambda \gtilde 10$TeV) put the lower bound as
$m_\phi \gtilde 100$ eV. Therefore the allowed region for the moduli
mass is obtained as $m_\phi \sim$ 100 eV--1 MeV.  We also find that if
we take $\VS \gtilde 10^8$ GeV, no allowed region exists because the
oscillating inflation moduli exceeds the cosmological constraints (See
Fig.\ref{fig:2}). Thus the oscillating inflation could give a
solution to the cosmological moduli problem if $\VS \ltilde 10^8$
GeV, and the gravitino mass lies in the region $m_{3/2} \simeq m_\phi
\sim$ 100 eV--1 MeV.

\section{Conclusion}

\label{sec:conclusion}

\indent

In this letter we have discussed the cosmological moduli problem in
the presence of the oscillating inflation caused by the flat direction
in the GMSB models. We have found that there are two types of the
coherent oscillation of the moduli, i.e. the big-bang moduli and the
oscillating inflation moduli and both coherent oscillations are
important. The former (latter) leads to the upper (lower) bound of
$\gm$. It has been shown that the oscillating inflation solves the
cosmological moduli problem if $\VS \ltilde 10^8$ GeV, and the
gravitino mass lies in the region $m_{3/2} \simeq m_\phi \sim$ 100
eV--1 MeV.

The allowed mass range of the gravitino comes from the requirement
that the present cosmic density of the moduli should not overclose the
universe for $m_{\phi} \ltilde 100$~keV and that the photon flux
produced by the moduli decay should not exceed the observed
x($\gamma$)-ray backgrounds for $m_{\phi} \gtilde 100$~keV. Thus, in
other words, the moduli with mass $m_{\phi} \sim 100$~eV-100~keV can
be dark matter of the universe. In particular, if the mass of the
moduli is $\sim 100$~keV, the x-ray flux from the moduli dark matter
will be detected in experiments with high energy resolution as
discussed in Ref.~\cite{AHKY2}.

In the present scenario, the primordial baryon asymmetry is also
diluted by the oscillating inflation.  Since the reheating temperature
is quite low ($\sim 10$~MeV), the electroweak or GUT baryogenesis does
not work at all. However, as shown in Ref.~\cite{GMM}, the
Affleck-Dine baryogenesis \cite{AD} can produce sufficient baryon
asymmetry in the present model if the moduli mass is small ($m_{\phi}
\ltilde 1$~MeV). We also show the constraint on $\Omega_{\phi}$ from
the baryon asymmetry in Fig.1 and Fig.2.

\subsection*{Acknowledgments}
MY is grateful to Professor K. Sato for his encouragement. This work
was partially supported by the Japanese Grant-in-Aid for Scientific
Research from the Monbusho, Nos.\ 10640250~(MK), 10-04558~(MY), and
``Priority Area: Supersymmetry and Unified Theory of Elementary
Particles(\#707)''(MK).

\newpage 

\def\NPB#1#2#3{Nucl. Phys. {\bf B#1}, #2 (19#3)} 
\def\PLB#1#2#3{Phys. Lett. {\bf B#1}, #2 (19#3)} 
\def\PLBold#1#2#3{Phys. Lett. {\bf#1B}, #2 (19#3)} 
\def\PRD#1#2#3{Phys. Rev. {\bf D#1}, #2 (19#3)}
\def\PRL#1#2#3{Phys. Rev. Lett. {\bf#1}, #2 (19#3)}
\def\PRT#1#2#3{Phys. Rep. {\bf#1}, #2 (19#3)} 
\def\ARAA#1#2#3{Ann. Rev. Astron. Astrophys. {\bf#1}, #2 (19#3)} 
\def\ARNP#1#2#3{Ann. Rev. Nucl. Part. Sci. {\bf#1}, #2 (19#3)} 
\def\MPL#1#2#3{Mod. Phys. Lett. {\bf #1}, #2 (19#3)} 
\def\PZC#1#2#3{Zeit. f\"ur Physik {\bfC#1}, #2 (19#3)} 
\def\APJ#1#2#3{Ap. J. {\bf #1}, #2 (19#3)}
\def\AP#1#2#3{{Ann. Phys. } {\bf #1}, #2 (19#3)}
\def\RMP#1#2#3{{Rev. Mod. Phys. } {\bf #1}, #2 (19#3)} 
\def\CMP#1#2#3{{Comm. Math. Phys. } {\bf #1}, #2 (19#3)} 
\def\PTP#1#2#3{{Prog. Thor. Phys. } {\bf #1}, #2 (19#3)}

\clearpage

\begin{figure}[htb]
  \centerline{\psfig{file=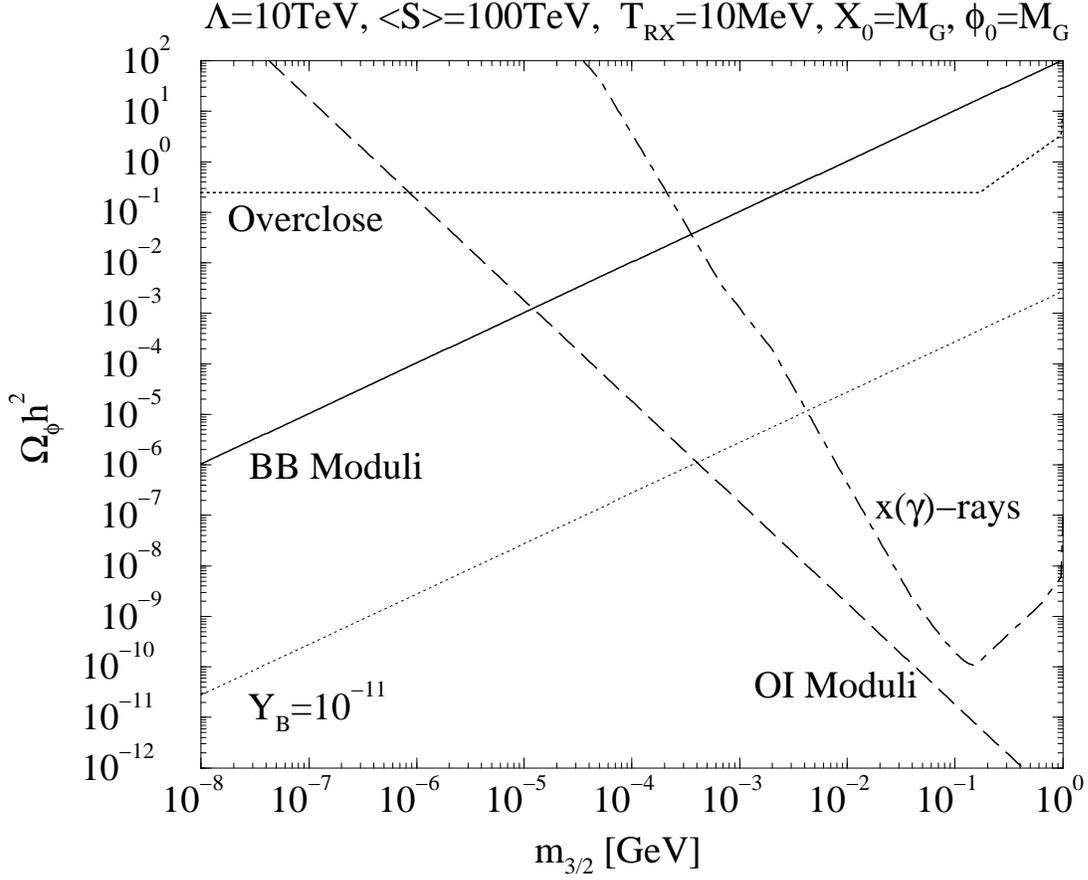,width=15cm}}
  \vspace{0.5cm}
  \caption{Abundances of the big-bang (BB) moduli (the solid line) and
  the oscillating inflation (OI) moduli (the long dashed line) for the
  case that $\Lambda = 10$ TeV, $\VS$ = 100 TeV, and $T_{RX} = 10$ MeV.
  We take $m_\phi = m_{3/2}$.  The upper bounds for the moduli
  abundance $\Omega_\phi$ from the present critical density (the
  dotted line) and the cosmic x($\gamma$) ray backgrounds (the
  dot-dashed line) are also shown.  We also show the lower bound from
  the present baryon-entropy ratio $Y_B = 10^{-11}$ (the thick dotted
  line).}
  \label{fig:1}
\end{figure}

\begin{figure}[htb]
  \centerline{\psfig{file=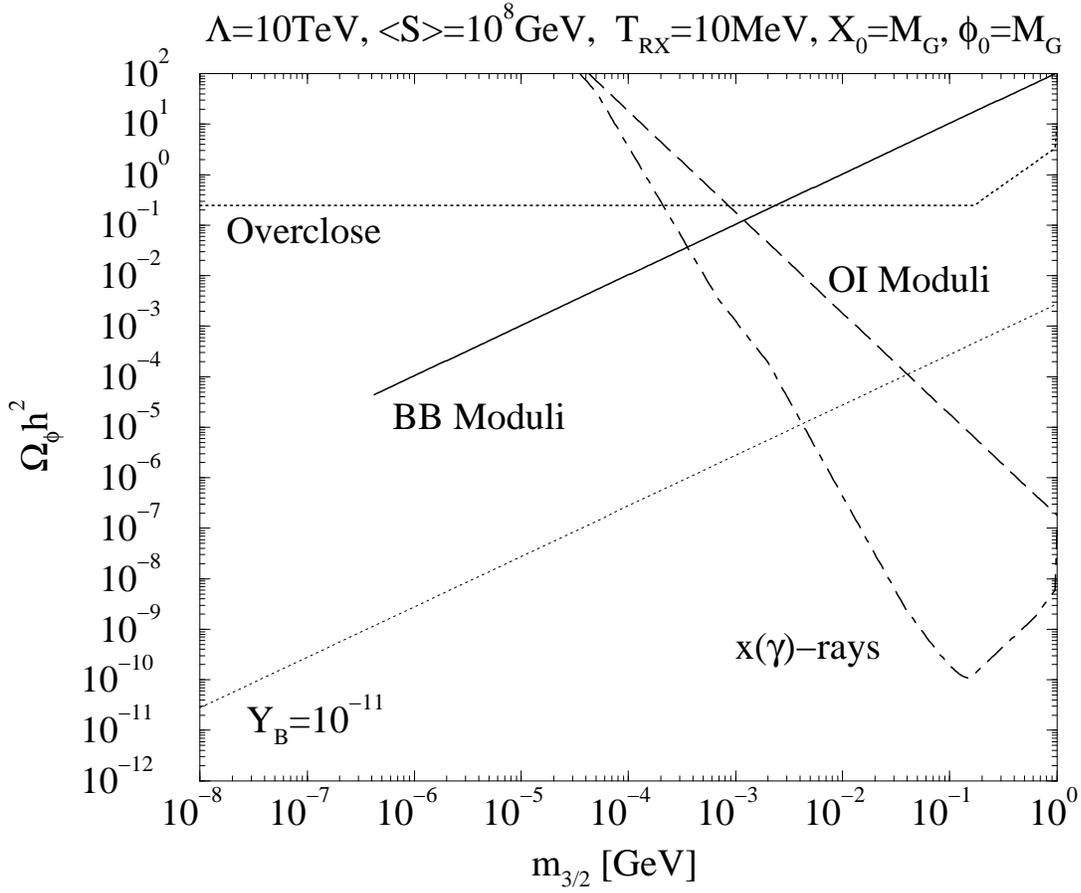,width=15cm}}
  \vspace{0.5cm}
  \caption{Same as Fig.1, except for $\VS=10^{8}$ GeV.
    In this figure we neglect moduli abundances for $m_{3/2} \lesssim
    100$ eV, since the condition (\ref{C-VS}) is broken. }
  \label{fig:2}
\end{figure}

\end{document}